\documentclass[aps, prl, twocolumn, letterpaper, nofootinbib, balancelastpage, longbibliography]{revtex4-1}

\usepackage{graphicx}
\usepackage{dcolumn}
\usepackage{amsmath, amsthm, mathrsfs, amsfonts, dsfont}
\usepackage{url}
\usepackage{xcolor}
\usepackage[colorlinks,citecolor=red,urlcolor=blue,linkcolor=red]{hyperref}
\usepackage{bm, bbold, xcolor, slashed}
\usepackage{tikz-feynman,contour}
\usepackage{verbatim}
\usepackage{subcaption}
\usepackage{easy-todo}
\usepackage{stmaryrd}
\usepackage{braket}
\usepackage{stackengine}
\usepackage{amssymb}
\usepackage{enumitem}

\newcommand{\bubble}{\rotatebox[origin=c]{-90}{$\between$}}

\begin{document}

\preprint{APS/123-QED}

\title{Scattering Gravitons off General Spinning Compact Objects to $\mathcal{O}(G^2S^4)$}
\author{Dogan Akpinar}
\email{dogan.akpinar@ed.ac.uk} 
\affiliation{Higgs Centre for Theoretical Physics, School of Physics and
Astronomy, University of Edinburgh, Edinburgh, EH9 3FD, UK} 

\date{\today}

\begin{abstract}
	We compute the classical one-loop gravitational Compton amplitude describing the scattering of a graviton off a massive spinning compact object at the second post-Minkowskian order, including terms through the quartic order in spin. Our analysis includes spin-induced finite-size effects up to the hexadecapolar order, and extends recent results obtained for minimal couplings at the quadratic order in spin. From the amplitude, we determine the scattering phase in momentum space, applicable in both the eikonal and wave regimes. In the eikonal limit, we then isolate the spin-independent contribution of the graviton field, explicitly linking it to the dynamics of a massless scalar probe in a Kerr background. This constitutes the first complete description of classical one-loop Compton scattering for generic spinning compact objects at the second post-Minkowskian and hexadecapolar orders.
\end{abstract}

\maketitle

\textbf{\textit{Introduction---}}Following the monumental detection of gravitational waves by the LIGO–Virgo–KAGRA collaboration~\cite{LIGOScientific:2016aoc,LIGOScientific:2017vwq}, an unprecedented observational window into the universe has opened, enabling high-precision studies of binary systems composed of massive, compact astrophysical objects like neutron stars and black holes. The advent of next-generation detectors demands corresponding theoretical progress---not only in precision, but also in our understanding of the fundamental aspects of classical two-body dynamics~\cite{Punturo:2010zz,LISA:2017pwj,Reitze:2019iox,Borhanian:2022czq,Purrer:2019jcp}.

Over the past several years, methods inspired by quantum field theory have driven remarkable advances in computing classical gravitational observables. Among these, the post-Minkowskian (PM) expansion---a perturbative expansion in Newton's constant $G$---has proven particularly powerful. For spinless binaries, substantial progress has been made using both on-shell scattering amplitude approaches~\cite{Cheung:2018wkq, Kosower:2018adc, Bern:2019nnu, Bern:2019crd, Cristofoli:2019neg, Bjerrum-Bohr:2019kec, Brandhuber:2021eyq, Bern:2021dqo, Bern:2021yeh, Damgaard:2023ttc, Kosmopoulos:2023bwc, Bern:2025zno} and worldline methods~\cite{Kalin:2020mvi, Kalin:2020fhe, Dlapa:2021vgp, kalin:2022hph, Dlapa:2022lmu, Dlapa:2023hsl, Dlapa:2021npj, Mogull:2020sak, Jakobsen:2021smu, Jakobsen:2022psy, Jakobsen:2023oow, Driesse:2024xad, Driesse:2024feo, He:2025how, Cheung:2023lnj, Cheung:2024byb}. Including spin effects, however, is essential for accurately modeling astrophysical systems, motivating extensive efforts in both amplitude- and worldline-based formulations~\cite{Porto:2005ac, Porto:2008jj, Porto:2016pyg, Bini:2017xzy, Bini:2018ywr, Vines:2017hyw, Vines:2018gqi, Guevara:2017csg, Guevara:2018wpp, Chung:2018kqs, Arkani-Hamed:2019ymq, Guevara:2019fsj, Chung:2019duq, Damgaard:2019lfh, Aoude:2020onz, Chung:2020rrz, Guevara:2020xjx, Bern:2020buy, Kosmopoulos:2021zoq, Chen:2021qkk,  FebresCordero:2022jts, Bern:2022kto, Bern:2023ity, Menezes:2022tcs, Riva:2022fru, Damgaard:2022jem, Aoude:2022thd, Aoude:2022trd, Bautista:2022wjf, Gonzo:2023goe, Aoude:2023vdk, Lindwasser:2023zwo, Brandhuber:2023hhl, DeAngelis:2023lvf, Aoude:2023dui, Bohnenblust:2023qmy, Gatica:2024mur, Cristofoli:2021jas, Luna:2023uwd, Gatica:2023iws, Liu:2021zxr, Jakobsen:2021lvp, Jakobsen:2021zvh, Jakobsen:2022fcj, Jakobsen:2022zsx, Jakobsen:2023ndj, Jakobsen:2023hig, Heissenberg:2023uvo, Lindwasser:2023dcv, Bautista:2023sdf, Cangemi:2023ysz, Brandhuber:2024bnz, Chen:2024mmm, Correia:2024jgr, Bhattacharyya:2024kxj, Alaverdian:2024spu, Brandhuber:2024qdn, Brandhuber:2024lgl, Akpinar:2024meg, Akpinar:2025bkt, Bohnenblust:2024hkw, Haddad:2024ebn, Bonocore:2024uxk, Akpinar:2025huz, Bohnenblust:2025gir, Aoki:2024boe, Hoogeveen:2025tew, Akpinar:2025tct, Aoki:2025ihc, Aoude:2025xxq}.

Within this context, a particularly significant quantity is the gravitational Compton amplitude, which describes the scattering of gravitons off a massive compact object. These amplitudes not only serve as fundamental building blocks in PM computations, but also encode information about the finite-size structure of generic compact bodies. In light of this, a natural direction is to systematically describe graviton scattering off generic massive compact objects through the inclusion of non-minimal couplings, which deform the minimal/Kerr dynamics. In particular, our focus will be on spin-induced non-minimal couplings, which have already been characterized through the quartic order in spin~\cite{Bern:2022kto, Haddad:2024ebn}. 

While a substantial body of existing work has concentrated on Kerr black holes at the first post-Minkowskian (1PM) order~\cite{Guevara:2018wpp, Chung:2018kqs, Bern:2022kto, Aoude:2022thd, Aoude:2023vdk, Bautista:2019tdr, Bautista:2022wjf, Arkani-Hamed:2017jhn, Falkowski:2020aso, Chiodaroli:2021eug, Cangemi:2022bew}, comparatively fewer studies have explored the dynamics of generic spinning compact objects~\cite{Haddad:2023ylx, Ben-Shahar:2023djm, Saketh:2022wap, Vazquez-Holm:2025ztz}. Moreover, the full classical gravitational Compton amplitude at the second post-Minkowskian (2PM) order has only very recently been computed for Schwarzschild black holes~\cite{Bjerrum-Bohr:2025bqg, Brunello:2025eso}\footnote{The full classical spinless one-loop Compton for electrodynamics has been computed in Ref~\cite{Brunello:2024ibk}.}, and for Kerr black holes through the quadratic order in spin~\cite{Akpinar:2025huz}. In fact, generic spinning bodies at this order have also previously been studied, but only within the eikonal limit~\cite{Chen:2022clh}. In this paper, we report the full classical one-loop gravitational Compton amplitude describing the scattering of a graviton off a generic massive spinning compact object at 2PM, $\mathcal{O}(G^2)$, including terms up to the quartic order in spin. This extends the recent 2PM results \cite{Bjerrum-Bohr:2025bqg, Akpinar:2025huz} to higher orders in spin, including spin-induced non-minimal couplings. From this result we extract the gauge-invariant scattering phase in momentum space---valid in the eikonal and wave regimes---whose position space representation in the eikonal regime serves as a generator of observables~\cite{Alessio:2025flu, Kim:2025hpn, Haddad:2025cmw, Akhtar:2025nmt, Ivanov:2025ozg}. Invoking the classical equivalence principle of general relativity, we extract the eikonal-regime graviton monopole contribution in position space and identify it with the dynamics of a massless scalar probe in a Kerr background for aligned-spin configurations. The monopole term---proportional to the squared Lorentz product of the graviton polarizations---encodes the spin-independent information of the massless field.

The organization of the paper is as follows. As our construction requires the regularization of loops, we  begin with an outline of the different dimensional regularization schemes considered throughout. Then, we overview the relevant tree-level amplitudes required for the generalized unitarity framework, focusing first on the Kerr dynamics. Following this, we describe the detailed computation of the classical one-loop Compton amplitude, providing a workflow agnostic as to whether the dynamics are minimal or non-minimal. In fact, several properties arising purely from quantum field theory considerations, such as Weinberg's infrared factorization, serve as strong consistency checks of our results. 
	Equipped with this knowledge, we obtain the classical one-loop Compton amplitude 
	that describes the scattering of a graviton off a Kerr black hole through the 
	quartic order in spin, thereby extending the results of Refs.~\cite{Bjerrum-Bohr:2025bqg, 
	Brunello:2025eso, Akpinar:2025huz}. In doing so, we find that the amplitude is dependent on 
	the dimensional regularization schemes that we consider in this paper 
	at the finite order of the dimensional regularization parameter, i.e.~at $
	\mathcal{O}(\epsilon^0)$, starting at the quadratic order in spin. Such 
	contributions, however, do not play a role in the scattering phase, which follows 
	by subtracting the tree iteration from the one-loop amplitude. Within the eikonal 
	regime, we find that the scattering phase for the graviton monopole contribution 
	exhibits the anticipated spin-shift symmetry through the quartic order in 
	spin~\cite{Chen:2022clh, Akpinar:2025huz}. Furthermore, in position space
	we find an exact agreement with the dynamics of a massless scalar probe in a Kerr background 
	for aligned-spin configurations. Incorporating spin-induced non-minimal 
	couplings~\cite{Haddad:2024ebn}, we further 
	calculate the non-minimal classical one-loop Compton amplitude through the quartic order in 
	spin. As before, we obtain the momentum space scattering phase, valid in both the eikonal 
	and wave regimes, as well as the eikonal-regime monopole contribution to the scattering phase in position space. 
	To conclude, we summarize our work and discuss future directions that build upon these results.

\textbf{\textit{Conventions---}}Throughout this paper we work in the negative signature with $\epsilon_{0123} = 1$, $\kappa^2 = 32\pi G$ and $c=1$. We adopt the convention in which factors of $2\pi$ are hidden as $\hat{\delta}^{(D)}(\cdot) = (2\pi)^D \delta^{(D)}(\cdot)$ and $\hat{\mathrm{d}}^Dk = \mathrm{d}^Dk/(2\pi)^D$. Averaged index symmetrization and anti-symmetrization are denoted as $x^{(\mu}y^{\nu)}$ and $x^{[\mu}y^{\nu]}$, respectively. All external legs in amplitudes are taken to be incoming unless stated otherwise. Finally, to regularize loop divergences we use dimensional regularization, in which the dimensions are taken to be $D = 4-2\epsilon$. 

\textbf{\textit{A note on regularization schemes---}}Before diving into the details of the classical one-loop Compton amplitude, it is important to discuss the regularization schemes employed in this work. We consider three schemes:	
	\begin{itemize}[leftmargin=0.4cm]
		\item \textit{'t Hooft-Veltman~(THV)}: external states are taken to be in four dimensions, while internal states and loop momenta are in $D = 4 - 2\epsilon$ dimensions~\cite{Bern:1991aq}. 
		\item \textit{Dimensional reduction~(DR)}: both external and internal states are in four dimensions, while loop momenta are in $D = 4 - 2\epsilon$ dimensions~\cite{Siegel:1979wq}.
		\item \textit{Conventional dimensional regularization~(CDR)}: all dimensional dependence is in $D = 4 - 2\epsilon$ dimensions.
	\end{itemize}
	The THV and DR schemes produce identical results for the classical 
	one-loop Compton amplitude, while the CDR scheme differs starting at the 
	$\epsilon^0$ order for quadratic and higher orders in spin. This difference stems 
	from explicit factors of $D$ appearing in contractions of the spin tensor in tree-
	level amplitudes. We will revisit this point later.

\textbf{\textit{Tree-level ingredients---}}Our goal is to construct the one-loop Compton amplitude using a generalized unitarity framework~\cite{Bern:1994zx, Bern:1994cg, Bern:1997sc, Britto:2004nc}. The advantage of such an approach is that the integrand of loop-level amplitudes are entirely determined---up to contact terms---directly from products of tree-level amplitudes. As a result, tree-level amplitudes serve as the fundamental building blocks in this program. In practice we work with heavy-mass expanded tree-level amplitudes to streamline the extraction of classical information~\cite{Damgaard:2019lfh, Aoude:2020onz, Brandhuber:2021eyq}.

Obtaining the desired classical one-loop Compton requires the classical, gauge-invariant three-point and tree-level Compton amplitudes through the quartic order in spin. The former is readily available within the literature in an exponentiated form~\cite{Guevara:2018wpp,Chung:2018kqs,Arkani-Hamed:2019ymq}, which we can be expanded to the desired order using the three-point kinematics~\cite{Bautista:2019tdr}
\begin{align}
	\mathcal{M}_{3,\mathrm{cl}}^{(0)} = - & \kappa m^2 (v \cdot \varepsilon)\Big[(v \cdot \varepsilon) + i(\varepsilon \cdot S \cdot k) \label{eq:3ptMinimal}\\
	& \hspace{-1cm}- \frac{1}{2!}(k \cdot S \cdot S \cdot k) + \frac{i}{3!}(\varepsilon \cdot S \cdot k)(k \cdot S \cdot S \cdot k) \nonumber \\
	& \hspace{-1cm} + \frac{1}{4!}(k \cdot S \cdot S \cdot k)^2 + \mathcal{O}(S^5)\Big]\nonumber \;.
\end{align}
Here the massive particle of mass $m$ has classical velocity $v^\mu$, while the graviton has momentum $k^\mu$ and polarization tensor $\varepsilon^{\mu \nu} = \varepsilon^\mu \varepsilon^\nu$. The spin of the massive particle is encoded in the mass-normalized spin tensor $S^{\mu \nu}$, which is related to the Pauli-Lubanski pseudo-vector $a^\mu$ via
\begin{equation}
	S^{\mu \nu} = \epsilon^{\mu \nu \rho \sigma}v_{\rho}a_{\sigma}\;, \quad a^{\mu} = \frac{1}{2}\epsilon^{\mu \nu \rho \sigma} v_{\nu} S_{\rho \sigma}\;.
\end{equation}
Crucially, the three-point kinematics in four-dimensions allows for the simplification $(k \cdot S \cdot S \cdot k) = - (k \cdot a)^2$.

The four-point classical tree-level Compton amplitude can be expressed compactly as~\cite{DeAngelis:2023lvf}
\begin{align}
    \mathcal{M}^{(0)}_{4,\mathrm{cl}} = & \frac{\kappa^2 m^2}{8(k_1 \cdot k_2)(v \cdot k_1)^2}\Big[\mathcal{J}_{0} \sum_{n = 0}^2  \mathcal{J}_{n} \label{eq:4ptMinimal}\\
    & \hspace{2cm} +\frac{\mathcal{J}_{1} \mathcal{J}_{2}}{3} + \frac{\mathcal{J}_{2}^2}{6} + \mathcal{O}(S^5) \Big] \nonumber\;,
\end{align}
together with
\begin{align}
    \mathcal{J}_{0} & = - 2 v \cdot F_1 \cdot F_2 \cdot v\;, \\[0.1cm]
    2i\mathcal{J}_{1} & = -2(k_1 \cdot F_2 \cdot v)\operatorname{Tr}(S \cdot F_1) \\
    & \hspace{1cm} + \operatorname{Tr}(S \cdot F_1 \cdot F_2)(k_1 - k_2)\cdot v + (1 \leftrightarrow 2)\;, \nonumber\\
    2\mathcal{J}_{2} & = (k_1 \cdot k_2)[\operatorname{Tr}(S \cdot S)(\operatorname{Tr}(F_1 \cdot F_2) +\mathcal{J}_{0})\\
    & \hspace{1cm} - 4 \operatorname{Tr}(S \cdot S \cdot F_1 \cdot F_2)]  \nonumber \\
    & \hspace{1cm} - \mathcal{J}_{0}(k_1 + k_2) \cdot S \cdot S \cdot (k_1 + k_2)\;, \nonumber
\end{align}
written in terms of the field strength $F_{i}^{\mu \nu} = 2k_i^{[\mu}\varepsilon_i^{\nu]}$ associated to each graviton leg. Here the trace products are defined as, for example, $\operatorname{Tr}(A \cdot B) = A^{\alpha \mu} B_{\mu \alpha}$ for any tensor $A$ and $B$. In addition to these amplitudes, we require the pure graviton four-point tree-level amplitude~\cite{Holstein:2006bh}, denoted as $\mathcal{M}_{h, 4}$. This will enter the generalized unitarity cuts used to assemble the classical one-loop Compton amplitude in the subsequent section.

It should be noted that the tree-level amplitudes discussed in this section are manifestly $D$-dimensional, since they are written in terms of spin tensors. Thus, strings of spin tensor contractions---particularly at the quadratic in spin and beyond---produce explicit factors of $D$ in the amplitude. These factors have consequences when choosing the dimensional regularization scheme, which we will outline when presenting results.
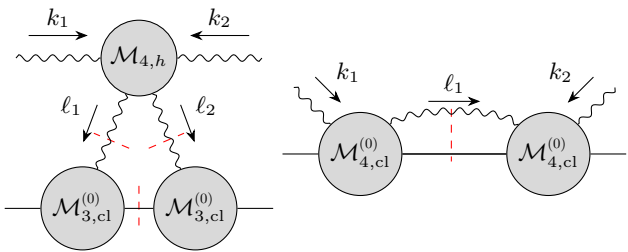
\begin{figure}[t]
    \centering
    \hspace*{-1.7cm}
    \begin{subfigure}[t]{0.2\textwidth}
    \centering
    \begin{tikzpicture}[baseline=(current bounding box.center)]
      \begin{feynman}[every blob={/tikz/fill=gray!30,/tikz/inner sep=2pt}]
         \vertex[blob] (b1) at (-0.75,-1) {$\mathcal{M}^{\scalebox{.6}{(0)}}_{3,\mathrm{cl}}$};
         \vertex[blob] (b2) at (0.75,-1) {$\mathcal{M}^{\scalebox{.6}{(0)}}_{3,\mathrm{cl}}$};
         \vertex[blob] (b3) at (0,1) {$\mathcal{M}_{4,h}$};
         \vertex (l1) at (-1.75,1) {};
         \vertex (l2) at (-1.9,-1) {};
         \vertex (r1) at (1.75,1) {};
         \vertex (r2) at (1.9,-1) {};
         \draw [dashed, color = red] ($(0,-0.7)$) -- ($(0,-1.3)$);
         \draw [dashed, color = red] ($(0.13,-0.2)$) -- ($(0.6,0)$);
         \draw [dashed, color = red] ($(-0.13,-0.2)$) -- ($(-0.6,0)$);
         \diagram*{
         (b3) -- [boson, rmomentum'={$k_1$}] (l1),
         (b3) -- [boson, rmomentum={$k_2$}] (r1),
         (b1) -- [plain] (l2),
         (b2) -- [plain] (r2),
         (b1) -- [plain] (b2),
         (b1) -- [boson] (b3),
         (b2) -- [boson] (b3),
         (b3) -- [draw=none, text=black, momentum'={[arrow shorten=0.25] $\ell_1$}] (b1),
         (b3) -- [draw=none, text=black, momentum={[arrow shorten=0.25] $\ell_2$}] (b2),
         };
       \end{feynman}
    \end{tikzpicture}
    \end{subfigure}
    \begin{subfigure}[t]{0.2\textwidth}
    \centering
    \begin{tikzpicture}[baseline=(current bounding box.center)]
      \begin{feynman}[every blob={/tikz/fill=gray!30,/tikz/inner sep=2pt}]
         \vertex[blob] (b1) at (-0.75,-1) {$\mathcal{M}^{\scalebox{.6}{(0)}}_{4,\mathrm{cl}}$};
         \vertex[blob] (b2) at (1.75,-1) {$\mathcal{M}^{\scalebox{.6}{(0)}}_{4,\mathrm{cl}}$};
         \vertex (l1) at (-1.75,0) {};
         \vertex (l2) at (-1.9,-1) {};
         \vertex (r1) at (2.75,0) {};
         \vertex (r2) at (2.9,-1) {};
         \draw [dashed, color = red] ($(0.46,-0.4)$) -- ($(0.46,-1.1)$);
         \diagram*{
         (b1) -- [boson, rmomentum'={$k_1$}] (l1),
         (b2) -- [boson, rmomentum={$k_2$}] (r1),
         (b1) -- [plain] (l2),
         (b2) -- [plain] (r2),
         (b1) -- [plain] (b2),
         (b1) -- [boson, quarter left,looseness=0.5] (b2),
         (b2) -- [text=black, rmomentum'={[arrow shorten=0.25,arrow distance=0.7cm] $\ell_1$}] (b1),
         };
       \end{feynman}
    \end{tikzpicture}
    \end{subfigure}
    \caption{The t-channel (left) and s-channel (right) cut configurations contributing to the classical one-loop Compton amplitude. The solid and wavy lines denote the heavy massive spinning particle and graviton, respectively. The red dashed lines are unitarity cuts.}
    \label{fig: one-loop-diag}
\end{figure}
%

\textbf{\textit{The Classical Compton Amplitude---}}All relevant contributions to the classical one-loop Compton amplitude are fixed by the t-channel and s-channel cut configurations; see Fig.~\ref{fig: one-loop-diag}. Within the generalized unitarity framework, the product of tree-level amplitudes requires a summation over physical states that can cross the unitarity cut. For massive legs, no such summation arises since the heavy-mass expansion enforces that all massive matter propagators are on-shell. For the massless legs, however, one requires the $D$-dimensional transverse graviton projector
\begin{align}
	P^{\mu \nu \rho \sigma} = \frac{1}{2}\left(\eta^{\mu \rho} \eta^{\nu \sigma} + \eta^{\mu \sigma} \eta^{\nu \rho}\right) - \frac{1}{D-2}\eta^{\mu \nu}\eta^{\rho \sigma}\;,
\end{align}
which we express in terms of the flat metric. This form, as opposed to the form with transverse spin-$1$ projectors, is appropriate since the tree-level amplitudes obey the generalized Ward identities (GWI)~\cite{Kosmopoulos:2020pcd}.

Concretely, the amplitude takes the form
\begin{align}
        \label{eq:flat space-one-loop}
        & \hspace{0.5cm} \mathcal{M}^{(1)}_{4,\mathrm{cl}}  = 
        c_{\scriptscriptstyle\Box}\mathcal{I}_{\scriptscriptstyle\Box} + 
        c_{\scriptscriptstyle\bigtriangleup}\mathcal{I}_{\scriptscriptstyle\bigtriangleup}  + 
        c_{\scriptscriptstyle\bubble} \mathcal{I}_{\scriptscriptstyle\bubble} + \mathcal{O}(m^2)\,, 
\end{align}
where $(\mathcal{I}_{\scriptscriptstyle\Box}, \mathcal{I}_{\scriptscriptstyle\bigtriangleup}, \mathcal{I}_{\scriptscriptstyle\bubble})$ denote the cut box, cut triangle, and cut bubble master integrals\footnote{Here \textit{cut} refers to the massive propagator being on-shell.}. Importantly, the pair of coefficients $c_{\scriptscriptstyle\Box},c_{\scriptscriptstyle\bigtriangleup}$ and $c_{\scriptscriptstyle\Box},c_{\scriptscriptstyle\bubble}$ and are fixed by the t-channel and s-channel cuts respectively. Once the cuts are computed we project the result into the form of \eqref{eq:flat space-one-loop} using the \texttt{FIRE6} program~\cite{Smirnov:2019qkx}. To do so, we organize all contributions into the single integral family discussed in Ref.~\cite{Bjerrum-Bohr:2025bqg} alongside irreducible scalar products involving spin. Note that since $c_{\scriptscriptstyle\Box}$ is determined by both cut configurations, this provides a non-trivial consistency check of our construction.

The integrals appearing in \eqref{eq:flat space-one-loop} are readily evaluated in
$D = 4 - 2\epsilon$~\cite{Bjerrum-Bohr:2025bqg}
\begin{align}
    \mathcal{I}_{\scriptscriptstyle\Box} & = i \bar{\mu}^{2\epsilon} \int \hat{\mathrm{d}}^{D} \ell \frac{\delta(v \cdot \ell)}{\ell^2 (\ell - q)^2[(\ell+k_1)^2 + i 0] } \nonumber \\
    & = -\frac{i}{16 \pi^2 \omega |\boldsymbol{q}|^2} \Bigg[\frac{1}{\epsilon} - \operatorname{log}\frac{|\boldsymbol{q}|^2}{\mu^{2}}\Bigg] + \mathcal{O}(\epsilon)\,, \nonumber \\[0.25cm]
   \mathcal{I}_{\scriptscriptstyle\bigtriangleup}  & =  \bar{\mu}
    ^{2\epsilon} \int \hat{\mathrm{d}}^{D} \ell \,\frac{\delta(v \cdot \ell)}{ \ell^2 (\ell-q)^2} = \frac{1}{16 \pi |\boldsymbol{q}|} + \mathcal{O}(\epsilon)\,,  \label{eq:master_heavy_mass} \\[0.25cm]
     \mathcal{I}_{\scriptscriptstyle\bubble} & =  i \bar{\mu}^{2\epsilon} \int \hat{\mathrm{d}}^{D} \ell\, \,\frac{\delta\left(v \cdot \ell \right)}{(\ell + k_1)^2+ i 0} \nonumber \\
    & = \frac{i \omega}{8\pi^2}\Big[1 + \epsilon\Big(i\pi + 2 - \operatorname{log}\frac{4 \omega^2}{\mu^2}\Big)\Big] + \mathcal{O}(\epsilon^2)\,, \nonumber
\end{align}
where $q = k_1 + k_2$, $q^2 = - |\boldsymbol{q}|^2$, and $\bar{\mu}^2 = \mu^2 e^{\gamma_{\text{E}}} /4\pi$ defines the usual modified subtraction scheme when performing dimensional regularization. Substituting the master integral values and discarding higher-order $\mathcal{O}(\epsilon)$ terms yields the final classical one-loop Compton amplitude:
\begin{align}
	\mathcal{M}^{(1)}_{4,\mathrm{cl}} = \frac{d_{\mathrm{IR}}}{\epsilon} & + d_1 \operatorname{log} \frac{4 \omega^2}{\mu^2} + d_2 \operatorname{log} \frac{\boldsymbol{q}^2}{\mu^2} \label{eq:oneLoop_Final}\\
	& \hspace{2cm} + d_{\mathrm{Im}}i + \frac{d_{\boldsymbol{q}}}{|\boldsymbol{q}|} + d_{\mathrm{R}} \nonumber \;,
\end{align}
where $\omega = v \cdot k_1 = - v \cdot k_2$ is the energy of the external gravitons and $v \cdot q = 0$. In this form, we first note that the amplitude exhibits the expected infrared divergence. Moreover, the coefficients appearing in \eqref{eq:oneLoop_Final} satisfy several important relations~\cite{Bjerrum-Bohr:2025bqg}
\begin{equation}
	d_{\mathrm{IR}} + d_1 + d_2 = 0\;, \ d_{\mathrm{R}} + i \pi d_1 = 0\;. 
\end{equation}
A further non-trivial check is provided by the Weinberg infrared factorization theorem, which states that the coefficient of the infrared divergence must be proportional to the tree-level amplitude~\cite{Weinberg:1965nx}; see also Refs.~\cite{Herderschee:2023fxh,Georgoudis:2023lgf,Brandhuber:2023hhy,Elkhidir:2023dco,Brunello:2025eso} for discussions of the five-point amplitude relevant for waveform computations. Explicitly, we will find the following relation through the quartic order in spin: 
\begin{equation}
	d_{\mathrm{IR}} = - \frac{i \kappa^2 m \omega}{16 \pi}\mathcal{M}^{(0)}_{4,\mathrm{cl}}\;.
	\label{eq:IRFactorization}
\end{equation}
For convenience, we further introduce the dimensionless parameter 
$y = - q^2 / 4\omega^2$~\cite{Bjerrum-Bohr:2025bqg}, which will streamline the 
presentation of the result.

\textbf{\textit{Kerr dynamics---}}Using the tree-level amplitudes in \eqref{eq:3ptMinimal} and \eqref{eq:4ptMinimal} as our building blocks, we now construct the classical one-loop Compton amplitude with minimal couplings. This amplitude is expected to describe the scattering of a graviton off a Kerr black hole; although this still remains to be confirmed by matching to black hole perturbation theory (BHPT)~\cite{Bautista:2021wfy,Bautista:2022wjf}.

Since the spin-dependent structures through the quartic order in spin become increasingly intricate, it is indeed important to emphasize that non-trivial four-dimensional identities relate otherwise distinct structures. Before selecting a minimal basis for these structures, we first adopt a symmetric kinematic parametrization for the external graviton momenta:
\begin{equation}
	k_1 = k + \frac{q}{2}\;, \quad k_2 = -k + \frac{q}{2}\;,
	\label{eq: avg_param}
\end{equation}
such that
\begin{equation}
	k\cdot q = 0\;, \quad 2k \cdot \varepsilon_1 = - q \cdot  \varepsilon_1\;, \quad 2k \cdot \varepsilon_2 = q \cdot  \varepsilon_2\;.
\end{equation}
Then, choosing the basis $\{v^\mu, k^\mu, q^\mu, n^\mu= \epsilon^{\mu \nu \rho \sigma} q_{\nu}v_{\rho}k_\sigma\}$, we proceed by decomposing all remaining vectors onto this basis; see the supplemental material for more details. The coefficients of \eqref{eq:oneLoop_Final} for the classical minimally coupled one-loop Compton amplitude through the quartic order in spin are provided in the ancillary file \texttt{ComptonAmpsAndObs.m}. These coefficients indeed satisfy the infrared factorization, the expected consistency relations, as well as gauge invariance through the quartic order in spin.

At this stage, we should discuss how the dimensional regularization scheme affects \eqref{eq:oneLoop_Final}. The first point is that $c_{\scriptscriptstyle\Box}$ and $c_{\scriptscriptstyle\bubble}$ are sensitive to the dimensional regularization scheme choice at the subleading $\epsilon$ order, that is at $\mathcal{O}(\epsilon)$ and $\mathcal{O}(\epsilon^0)$ respectively. On the one hand, substituting the master integral values reveals that these differences do not appear in $d_{\mathrm{IR}}, d_1, d_2, d_{\boldsymbol{q}}$ and $d_{\mathrm{R}}$ for all three schemes; these terms receive contributions only from the leading $\epsilon$ expansion of the master coefficients, which are dimensional regularization scheme independent. On the other hand, $d_{\mathrm{Im}}$ is directly affected by the subleading $\epsilon$. Interestingly, it turns out that any differences at the subleading order in $\epsilon$ for the THV and DR schemes cancel to give an identical result. However, the CDR scheme gives a different coefficient altogether starting at the quadratic order in spin. Although one might initially be skeptical of this scheme dependence, it is important to note that $d_{\mathrm{IR}}, d_1, d_2, d_{\boldsymbol{q}}$ and $d_{\mathrm{R}}$ follow from the leading $\epsilon$ expansions of the master coefficients, yielding the consistency relations: $d_{\mathrm{IR}} + d_1 + d_2 = 0$ and  $d_\mathrm{R} + i \pi d_1 = 0$. As such, $d_{\mathrm{Im}}$ is the only contribution which is unconstrained, so it is somewhat unsurprising that it is sensitive to regularization-scheme-dependent effects. 

In light of this observation, an important question becomes: are physical observables affected by regularization scheme dependences? It has recently been pointed out in Ref.~\cite{Bini:2024rsy} that the spinless waveform at one-loop is dimensional regularization scheme dependent. As our discussion has revealed that the cut box and cut bubble master coefficients are dimensional regularization scheme dependent in the presence of spin, paired with the fact that the one-loop Compton amplitude plays a pivotal role in loop-level computations of the waveform, this highlights one source of scheme dependence that must be tracked consistently. It would be interesting to see exactly how this dimensional regularization scheme dependence affects the one-loop waveform, but we leave such a tantalizing endeavor to future work. If we restrict ourselves to non-radiative observables, however, then any scheme dependence is expected to drop out. This is the focus of the next section.

\textbf{\textit{Observables---}}As an application of our discussion, we now provide a connection with non-radiative observables. In particular, the scattering phase is related to the scattering amplitude through the exponential representation of the $S$-matrix~\cite{Damgaard:2021ipf, Gonzo:2024zxo, Kim:2024svw, Alessio:2022kwv, Kim:2025hpn, Akhtar:2025nmt, Ivanov:2025ozg, Caron-Huot:2025tlq}. In light of this, our starting point is the exponential form of the $S$-matrix
\begin{equation}
	S = 1 + i T = e^{i \Delta}\;,
	\label{eq: SMatrix_Exp}
\end{equation}
where $S$ is expressed in terms of the $T$-matrix, which is ultimately related to the scattering amplitude, whereas $\Delta$ will be related to the momentum space scattering phase. Within the context of perturbation theory, we proceed by expanding $T$ and $\Delta$ perturbatively in Newton's constant
\begin{equation}
	T = \sum_{n = 0}^{\infty} G^{n + 1} T^{(n)}\;, \quad \Delta = \sum_{n = 0}^{\infty} G^{n + 1} \Delta^{(n)}\;,
	\vspace{0.5em}
	\label{eq: TAndDelta_GExpansion}
\end{equation}
such that at 2PM \eqref{eq: SMatrix_Exp} gives
\begin{align}
	T^{(0)} = \Delta^{(0)}\;, \quad T^{(1)} = \Delta^{(1)} + \frac{i}{2}\Delta^{(0)}\Delta^{(0)}\;.
\end{align}
Using the unitarity of the $S$-matrix
\begin{equation}
	1 = (1 + iT)^\dagger (1 + iT) = 1 - iT^\dagger + i T + T^\dagger T\;,
\end{equation}
gives $T = T^\dagger + i T^\dagger T$. Inverting the relation for $T^{(1)}$ in favor of $\Delta^{(1)}$ yields
\begin{equation}
	\Delta^{(1)} = T^{(1)} - \frac{i}{2} (T^{(0)})^\dagger T^{(0)}\;,
\end{equation}
where we have used the relation following from unitarity. From this, we may acquire a form applicable to scattering amplitudes by taking an expectation value with respect to external states. Doing this explicitly involves inserting a complete set of states between $T^\dagger T$, resulting in a two-particle cut contribution. More specifically, this subtracts off the imaginary part of the one-loop amplitude coming from the cut box and cut bubble masters for even orders in spin, which leaves behind the real finite part. For odd orders in spin, however, the spin structures carry an extra factor of imaginary unit such that the real parts (which are imaginary for the even in spin orders) cancel to leave behind the finite imaginary part (which are real for the even in spin orders). This is essentially the optical theorem. Thus, we find that the expectation value of $\Delta$ at one-loop reads
\begin{equation}
	\langle \Delta^{(1)} \rangle = \frac{d_{\boldsymbol{q}}}{|\boldsymbol{q}|} + d_{\mathrm{R}}\;.
	\label{eq: ScatteringPhase_MomSpace}
\end{equation}
Consequently, the infrared divergence, the superclassical piece, and the dimensional regularization scheme dependence all cancel. It is important to stress that \eqref{eq: ScatteringPhase_MomSpace} is now interpreted as the scattering phase in momentum space, which is valid both in the eikonal and wave regimes. 

Through the classical equivalence principle in general relativity, it is expected that the monopole contribution of the graviton reproduces the dynamics of a massless scalar probe propagating in a Kerr background. In order to make a connection with this, we extract the $(\varepsilon_1 \cdot \varepsilon_2)^2$ contribution~\cite{Bjerrum-Bohr:2016hpa} and restrict ourselves to the eikonal---or, in other words, the geometric optics---regime
\begin{equation}
	|\boldsymbol{q}| \ll \omega \ll m\;,
\end{equation}
to find 
\begin{equation}
	\langle \Delta^{(1)} \rangle \Big \rvert_{(\varepsilon_1 \cdot \varepsilon_2)^2} = \frac{G^2 m^3}{|\boldsymbol{q}|}\sum_{j = 0}^{4}\alpha^{(j)}\;.
	\label{eq:one_loop_light_bending}
\end{equation}
In this limit, the finite part coming from the bubble $d_\mathrm{R}$ is quantum suppressed as compared to the triangle $d_{\boldsymbol{q}}$, such that only the latter contributes in the classical limit. Indeed the coefficients in \eqref{eq:one_loop_light_bending} exhibit spin-shift symmetry through the quartic order in spin\footnote{Note that spin-shift symmetry is not present beyond the eikonal limit, nor is it present beyond the monopole contribution of the graviton.}; as previously observed in Refs.~\cite{Chen:2022clh,Akpinar:2025huz}. Explicitly, the coefficients read:
\begin{widetext}
\begin{align}
	& \hspace{4cm} \alpha^{(0)} = 30 \pi^2 \omega^2\;, \quad \alpha^{(1)} = -40 i \pi^2 \omega (n \cdot a)\;, \nonumber \\[0.1cm]
	& \hspace{3.05cm} 4 \alpha^{(2)} = 130 \pi^2  |\boldsymbol{q}|^2 (k \cdot a)^2 + 95\pi^2\omega^2 \left[(q \cdot a)^2 + |\boldsymbol{q}|^2 a^2\right]\;, \noindent\\[0.1cm]
	& \hspace{3cm} \omega \alpha^{(3)} =  -16 i \pi^2 |\boldsymbol{q}|^2 (n \cdot a)(k \cdot a)^2 - 9 i \pi^2 \omega^2\left[(q \cdot a)^2 + |\boldsymbol{q}|^2 a^2\right]\;, \nonumber \\[0.1cm]
	& 96 \omega^2 \alpha^{(4)} = 512 \pi^2 |\boldsymbol{q}|^4 ( k \cdot a)^4 + 772 \pi^2 \omega^2 |\boldsymbol{q}|^2 (k \cdot a)^2 \left[(q \cdot a)^2 + |\boldsymbol{q}|^2 a^2\right] + 239 \pi^2 \omega^4 \left[(q \cdot a)^2 + |\boldsymbol{q}|^2 a^2\right]^2\;. \nonumber
\end{align}
\end{widetext}

The first step to compute observables is obtaining the scattering phase in position space. Focusing on the graviton monopole contribution, we perform the Fourier transform of \eqref{eq:one_loop_light_bending}
\begin{equation}
	\delta^{(1)} = \int \hat{\mathrm{d}}^4 q \ \hat{\delta}(2 m v \cdot q) \hat{\delta}( 2 k \cdot q) e^{i b \cdot q} \langle \Delta^{(1)} \rangle \Big \rvert_{(\varepsilon_1 \cdot \varepsilon_2)^2}\;,
\end{equation}
where $b$ is the impact parameter, which is conjugate to the momentum transfer $q$. As the rank of the integral we must evaluate grows with the power of spin, we will start with the scalar integral
\begin{align}
	I_\alpha & = \int \hat{\mathrm{d}}^4 q \ \hat{\delta}(2 m v \cdot q) \hat{\delta}( 2 k \cdot q) e^{i b \cdot q} (-q^2)^{-\alpha} \\
	& = \frac{1}{4m}\int \hat{\mathrm{d}}^4 q \ \hat{\delta}(v \cdot q) \hat{\delta}( k \cdot q) e^{i b \cdot q} (-q^2)^{-\alpha} \nonumber\;.
\end{align}
This can be performed by noticing that $v^\mu = (1,0,0,0)$ and choosing to align the graviton momentum with the $z$-axis: $k^\mu = (\omega, 0, 0, \omega)$~\cite{Cristofoli:2021vyo}. Although the parametrization in \eqref{eq: avg_param} enforces $4 k^2 = - q^2$, in the eikonal regime $|\boldsymbol{q}| \ll \omega$ we have $k ^2 = 0$. This yields~\cite{Maybee:2019jus, Herrmann:2021tct}
\begin{align}
	I_\alpha & = \frac{1}{4m \omega}\int \hat{\mathrm{d}}^4 q \ \hat{\delta}(q^0) \hat{\delta}( q^0 - q^3) e^{i b \cdot q} (-q^2)^{-\alpha} \\
	&  = \frac{1}{4m \omega}\int \hat{\mathrm{d}}^2 \boldsymbol{q} \ e^{-i \boldsymbol{b} \cdot \boldsymbol{q}} (\boldsymbol{q}^2)^{-\alpha} = \frac{1}{m\omega}\frac{\Gamma(1-\alpha)}{2^{2\alpha + 2} \pi \Gamma(\alpha)} \frac{1}{|b|^{2-2\alpha}}\;, \nonumber 
\end{align}
where we define $\boldsymbol{b}^2= -  b^2 \equiv |b|^2$. To evaluate higher rank integrals we simply perform differentiation with respect to $b^\mu$ since the scalar integral is solely a function of $b^2$:
\begin{align}
	I^{\mu_1 \ldots \mu_n}_\alpha (b^2) & = -i \frac{\partial }{\partial b^{\mu_n}}I^{\mu_1\ldots \mu_{n-1}}_\alpha (b^2)\;.
\end{align}
The relevant integrals to $n = 4$ are in the supplemental material, while the monopole contribution to the scattering phase in position space can be found in the ancillary file \texttt{ComptonAmpsAndObs.m}. Note that we do not include the full momentum space scattering phase (which is also valid in the wave regime), as the coefficients of the classical one-loop Compton amplitude are readily found in the file.

In the presence of spin the scattering process is non-planar, i.e.~it is not confined to a plane, meaning that the dynamics is parametrized by several angles. For our purposes, we are interested in computing the scattering on the equatorial plane, which we achieve with an aligned-spin configuration. The aligned-spin dynamics follows by aligning the spin vector with the angular momentum
\begin{align}
	& a \cdot k =a \cdot \hat{b} = 0\;, \quad a \cdot a = - |a|^2\;, \\
	& \hspace{1cm} \epsilon^{\mu \nu \rho \sigma} v_\nu k_\rho a_\sigma = \omega |a| \hat{b}^\mu\;, \nonumber
\end{align}
with normalized impact parameter $\hat{b}^\mu = b^\mu/|b|$. In this scenario, we find that the scattering phase takes the form
\begin{align}
	\hspace{-0.5cm} \frac{\delta^{(1)}}{G^2 \pi m^2 \omega} =  & \ \frac{15 \omega}{4J} - \frac{5 \omega^2}{J^2}|a| + \frac{95 \omega^3}{16 J^3}|a|^2 \\
	& \hspace{0.2cm} - \frac{27 \omega^4}{4 J^4}|a|^3 + \frac{239 \omega^5}{32 J^5}|a|^4 + \mathcal{O}(|a|^5)\;, \nonumber
\end{align}
in exact agreement with Ref.~\cite{Bautista:2021wfy}. The scattering angle follows from the derivative of the scattering phase with respect to the angular momentum in the asymptotic past $J = \omega |b|$:
\begin{equation}
	\chi = -\frac{\partial \delta}{\partial J} = -\frac{1}{\omega} \frac{b_\mu}{|b|} \frac{\partial \delta}{\partial b_\mu}\;.
	\label{eq: ScatteringAngle}
\end{equation}
Thus, we obtain
\begin{align}
	\hspace{-0.5cm} \frac{\chi^{(1)}}{G^2 \pi m^2 \omega} =  & \ \frac{15 }{4J^2} - \frac{10 \omega}{J^3}|a| + \frac{285 \omega^2}{16 J^4}|a|^2 \\
	& \hspace{0.2cm} - \frac{27 \omega^3}{J^5}|a|^3 + \frac{1195 \omega^4}{32 J^6}|a|^4 + \mathcal{O}(|a|^5)\;. \nonumber
\end{align}

\textbf{\textit{Beyond Kerr dynamics---}}To deform the classical one-loop Compton amplitude beyond the Kerr black hole case and thereby describe the dynamics of a generic spinning compact object, we must incorporate spin-induced non-minimal effects from the quadratic order in spin onward. Our starting point is the bosonic worldline model of Ref.~\cite{Haddad:2024ebn}, which includes such deformations through linear-curvature and leading higher-curvature couplings. Within this framework, the worldline action describing a massive spinning point particle with spin-induced non-minimal couplings reads
\begin{equation}
	S = - m \int d \tau \left(\frac{1}{2}g_{\mu \nu} \dot{x}^\mu \dot{x}^\nu - i \bar{\alpha}_b \frac{D \alpha^b}{d \tau} - \mathcal{L}_{\mathrm{nm}}^{\leq 4}\right)\;,
	\label{eq:BUSY_action}
\end{equation}
where $\tau$ is the proper time, $m$ is the mass and $x^\mu(\tau)$ is the particle trajectory. The spinning degrees of freedom are captured by a conjugate pair of commuting complex vectors $(\bar{\alpha}^b(\tau), \alpha^b(\tau))$---known as bosonic oscillators---that are defined in a locally flat spacetime\footnote{Global spacetime indices are denoted with Greek letters, whereas the local spacetime indices are denoted with Latin letters.}. These objects are embedded into the globally curved spacetime through the action of a vielbein $e^\mu_b(x)$. Indeed one obtains the non-mass-renormalized spin tensor from the bosonic oscillator as
\begin{equation}
	S^{\mu \nu} = - 2im \bar{\alpha}^{[\mu} \alpha^{\nu]}\;.
\end{equation}
The second term in \eqref{eq:BUSY_action} involves the spin connection $\omega_{\mu}^{~bc}$ in the definition of the derivative
\begin{equation}
	\frac{D \alpha^b}{d \tau} = \dot{\alpha}^b + \dot{x}^\mu \omega_{\mu}^{bc}\alpha_c\;.
\end{equation}
The important part that describes the non-minimal coupling information is contained in $\mathcal{L}_{\mathrm{nm}}^{\leq 4}$. Here the superscript $\leq 4$ indicates that we only concern ourselves with spin-induced non-minimal couplings through the quartic order in spin. In this case there are six operators, which are organized into two types of contributions~\cite{Haddad:2024ebn,Haddad:2025cmw}
\begin{equation}
	\mathcal{L}_{\mathrm{nm}}^{\leq 4} = \mathcal{L}_{\mathrm{nm}}^{(a)} + \mathcal{L}_{\mathrm{nm}}^{(Z)}\;.
\end{equation}
The right-hand-side is understood to solely contain the non-minimal coupling terms appearing through the quartic order in spin. More explicitly, these terms read
\begin{align}
	& \hspace{-0.05cm} \mathcal{L}_{\mathrm{nm}}^{(a)} = \frac{ C_{\mathrm{ES}^2}}{2!}\mathcal{E}_{\mathfrak{a} \mathfrak{a}} -\frac{ i C_{\mathrm{BS}^3}}{3!}\nabla_\mathfrak{a} \mathcal{B}_{\mathfrak{a}\mathfrak{a}} + \frac{ C_{\mathrm{ES}^4}}{4!}(\nabla_\mathfrak{a})^2\mathcal{E}_{\mathfrak{a} \mathfrak{a}}\;, \\
	& \hspace{-0.05cm} \mathcal{L}_{\mathrm{nm}}^{(Z)} = \frac{C_{Z^2}}{2}\mathcal{B}_{\mathfrak{a} \mathfrak{z}} - \frac{\nabla_\mathfrak{z}}{|\dot{x}|}\Big( \frac{i C_{Z^3}}{3!} \mathcal{E}_{\mathfrak{a}\mathfrak{a}} + \frac{C_{Z^4}}{4!}\nabla_\mathfrak{a} \mathcal{B}_{\mathfrak{a}\mathfrak{a}}\Big)\;,
\end{align}
where $\mathcal{E}_{\mu\nu}$ and $\mathcal{B}_{\mu\nu}$ are electric/magnetic curvature tensors 
evaluated on the background trajectory, and $\nabla_\mu$ is the usual covariant derivative. They are contracted with
\begin{equation}
	\mathfrak{a}^\mu = \frac{1}{|\dot{x}|} \epsilon^{\mu \dot{x} \bar{\alpha} \alpha}\;, \quad \mathfrak{z}^\mu = \frac{2 \bar{\alpha}^{[\mu} \alpha^{\nu]}\dot{x}_\nu}{|\dot{x}|}\;,
\end{equation}
using Schoonschip notation, e.g., $\nabla_\mathfrak{a} = \mathfrak{a}^\mu \nabla_\mu$~\cite{Strubbe:1974vj, Veltman:1991xb}. Here $C_{\mathrm{ES}^n}$ and $C_{\mathrm{BS}^n}$ are electric-type and magnetic-type Wilson coefficients, while $C_{Z^n}$ are constrained to take the following values~\cite{Haddad:2024ebn}
\begin{equation}
	C_{Z^2} = 2\;, \quad C_{Z^3} = 3 C_{\mathrm{ES}^2}\; \quad C_{Z^4} = 4 C_{\mathrm{BS}^3}\;,
\end{equation}	
such that the dynamics preserves the spin-supplementary condition $v_\mu S^{\mu \nu} = 0$ where $v^\mu = \dot{x}^\mu$. One can then go ahead and derive Feynman rules via a weak-field expansion $g_{\mu \nu} = \eta_{\mu \nu} + \kappa h_{\mu \nu}$ where $h_{\mu \nu}$ is graviton field. At the same time, one must perturb the trajectory and bosonic oscillators around their background values; see Ref.~\cite{Haddad:2024ebn} for more details. Importantly, we should note that the Feynman rules obtained from this model is valid in four-dimensions only, meaning that for the non-minimal computation we employ the THV and DR schemes. 

We then proceed with computing the tree-level three-point and four-point amplitudes involving non-minimal couplings. It is important to note the difference in convention between the worldline and amplitude approaches, whereby the conventional amplitude is obtained from the on-shell worldline amplitude multiplied by an additional factor of $2m$~\cite{Haddad:2025cmw}. In what follows we will always convert to the conventional amplitude form to be consistent with previous discussions. To this end, the non-minimal classical three-point amplitude follows from the single graviton emission from the worldline evaluated on-shell
\begin{align}
	\mathcal{M}_{3,\mathrm{nm}}^{(0)} = - & \kappa m^2 (v \cdot \varepsilon)\Big[(v \cdot \varepsilon) + i(\varepsilon \cdot S \cdot k) \label{eq:3ptNonMinimal}\\
	& \hspace{-1cm} + \frac{C_{\mathrm{ES}^2}}{2!}(k \cdot a)^2 + \frac{iC_{\mathrm{BS}^3}}{3!}(\varepsilon \cdot S \cdot k)(k \cdot a)^2 \nonumber \\
	& \hspace{-1cm} + \frac{C_{\mathrm{ES}^4}}{4!}(k \cdot a)^4 + \mathcal{O}(S^5)\Big] \nonumber\;.
\end{align}
Obtaining the non-minimal classical four-point Compton amplitude follows by constructing all possible diagrams with two graviton emissions from the worldline. Specifically, we need the diagrams with no external deflections, and, at most, a single internal deflection. There are three contributions to this amplitude: the $t$-channel diagram, which comes from the contraction of the single graviton emission from the worldline with the three-graviton vertex; the diagrams involving propagating worldline and oscillator deflections; and contact terms coming from the double graviton emission from the worldline. 

One must actually take care when using the tree-level Compton amplitude acquired here within the generalized unitarity framework, since Ref.~\cite{Haddad:2025cmw} pointed out that the cubic in spin and quartic in spin contributions do not satisfy GWI. Here we find that only the cubic in spin terms do not satisfy this property, unless we impose four-dimensional identities; practically speaking, this is done by substituting a general frame valid in four-dimensions to make this property manifest. It seems that the cubic in spin structures also do not satisfy the expected Ward identity (WI) unless the four-dimensional identities are imposed. Of course, if we wish to use this amplitude in the generalized unitarity framework, GWI (or even WI) must be manifest. The standard way to remedy this would be to impose all four-dimensional identities by choosing a basis. In practice, however, this is complicated by the external graviton legs, which lead to many different Levi-Civita structures and momentum dependent denominators. By diagnosing the amplitude in detail, we find that contributions proportional to $C_{\mathrm{BS^3}}$ break this property. To circumvent this problem we 
	simply add a Gram determinant---which vanishes in four-dimensions---to restore the expected 
	property, without affecting the amplitude itself, at least in four-dimensions\footnote{This 
	was determined by simply taking the difference of the non-minimal tree-level amplitude with 
	Kerr values and minimal tree-level amplitude.}. One can think of this as simply adding a zero 
	in four-dimensions. Once this has been done, the WI and GWI are indeed manifestly restored. 
	For total transparency, both versions of the amplitude can be found in the ancillary file 
	\texttt{ComptonAmpsAndObs.m}.

As a nice check, in the supplemental material we evaluate this amplitude in an arbitrary four-dimensional basis and for Kerr values $C_{\mathrm{ES}^n} = C_{\mathrm{BS}^n} = 1$. In this case, we notice that the amplitude takes a compact form wherein the $\mathcal{O}(a^n)$ terms are proportional to the spinless contribution~\cite{Bautista:2021wfy, Bautista:2022wjf}.

Using these non-minimal tree-level amplitudes as the ingredients of our generalized unitarity pipeline yields the non-minimal classical one-loop Compton amplitude. Importantly, we find that the usual Kerr values ($C_{\mathrm{ES}^n} = C_{\mathrm{BS}^n} = 1$) reproduce the minimally coupled amplitude. A crucial and immediate check is the agreement of the cut box coefficient obtained via the $t$- and $s$-channel cut configurations. In fact, this is a non-trivial check since the non-minimal information is encoded very differently in the three- and four-point tree-level amplitudes. Moreover, we find that the amplitude satisfies the expected infrared factorization \eqref{eq:IRFactorization}, relation between coefficients, and gauge invariance. On the note of scheme dependence, we indeed find that the cut box and bubble master values are scheme dependent, but this dependence drops out after substituting masters, since we only use the THV and DR schemes. Finally, we compute the momentum space scattering phase and extract the monopole contribution to the graviton in the eikonal regime, which we Fourier transform into position space. As expected, we find that spin-shift symmetry is only satisfied by the eikonal-regime monopole contribution in momentum space for the Kerr values of the Wilson coefficients~\cite{Aoude:2020ygw,Kosmopoulos:2021zoq,Chen:2022clh,Akpinar:2025tct}. Finally, we compute the aligned-spin scattering angle.

As before, the non-minimal classical one-loop Compton amplitude through the quartic order in spin, as well as the monopole contribution to the scattering phase in position space and scattering angle, can be found in \texttt{ComptonAmpsAndObs.m}.

\textbf{\textit{Conclusion---}}In this paper, we computed the classical one-loop Compton amplitude describing the scattering of a graviton off a generic massive spinning compact object through the quartic order in spin. This result provides the complete 2PM contribution to the dynamics of generic spinning bodies, extending previous analyses that were limited to minimal couplings, lower spin orders or the eikonal limit.

Our construction here incorporates both minimal and spin-induced non-minimal interactions, thereby capturing spin-induced finite-size effects associated with generic compact objects beyond the Kerr solution. The resulting amplitude satisfies several non-trivial consistency checks, including the Weinberg infrared factorization, expected relations among coefficients, as well as gauge invariance. Interestingly, we found that the one-loop Compton amplitude seems to be dimensional regularization scheme dependent in the presence of spin, which we anticipate to be a source of scheme dependence appearing in waveform computations~\cite{Bini:2024rsy}. The one-loop Compton amplitude is indeed an essential ingredient in, for example, computations of the loop-level waveform~\cite{Herderschee:2023fxh,Georgoudis:2023lgf,Brandhuber:2023hhy,Elkhidir:2023dco, Bohnenblust:2023qmy, Bohnenblust:2025gir, Brunello:2024ibk, Brunello:2025eso}. 

To study non-radiative observables we extracted the scattering phase, which we found to be independent of the infrared divergence, the superclassical contributions, and the dimensional regularization scheme. In fact, this constitutes the momentum space scattering phase which is valid in the eikonal and wave regimes. By analyzing the monopole contribution of the graviton in the eikonal regime in position space---with and without non-minimal couplings---we found exact agreement with the dynamics of a massless scalar probe in a Kerr background, for Kerr values of Wilson coefficients. As anticipated, we found that this contribution to the scattering phase in momentum space exhibits spin-shift symmetry for Kerr values of the Wilson coefficients~\cite{Chen:2022clh,Akpinar:2025huz}. Finally, from the position space scattering phase we further extracted the aligned-spin scattering angle. 

The results here not only deepen our understanding of the structure of classical gravitational scattering, but also provide essential input for precision modeling of spinning binary dynamics and gravitational-wave observables. In light of all of this, future directions include extending this analysis to higher PM orders, incorporating dissipative effects, and including higher orders in spin. An essential endeavor would be a detailed matching with BHPT to confirm that minimal coupling really captures the dynamics of Kerr beyond the 1PM order. This thereby paves the way for a more complete description of the interactions of compact object within the modern amplitude/worldline framework.

\textbf{\textit{Acknowledgements---}}I thank Rafael Aoude, Fabian Bautista, Graham R. Brown, Giacomo Brunello, Asaad Elkhidir, Riccardo Gonzo, Matteo Sergola and Mao Zeng for many insightful discussions and comments on the draft.
I am especially grateful to Kays Haddad for sharing private files for the Feynman rules of Ref.~\cite{Haddad:2024ebn} and for many insightful discussions.
I have extensively used the Mathematica package {\tt FeynCalc}
\cite{Shtabovenko:2023idz} for some of these calculations.
D.A.\ is supported by an STFC studentship.

\bibliography{oneLoopCompton}
\newpage
\appendix
\onecolumngrid
\vspace{0.8cm}
\begin{center}
\textbf{Supplemental Material}
\end{center}

\section{Decomposing onto a convenient basis}
\noindent Having chosen the basis discussed in the main text, we proceed by decomposing all other vectors as follows:
\begin{align}
	& \varepsilon_1^\mu = \lambda_1 v^\mu + \lambda_2 k^\mu + \lambda_3 q^\mu + \lambda_4 n^\mu\;, \label{eq:ep1_decomp}\\
	& \varepsilon_2^\mu = \xi_1 v^\mu + \xi_2 k^\mu + \xi_3 q^\mu + \xi_4 n^\mu\;, \label{eq:ep2_decomp}\\
	& a^\mu = \zeta_1 v^\mu + \zeta_2 k^\mu + \zeta_3 q^\mu + \zeta_4 n^\mu\;. \label{eq:a_decomp}
\end{align}
To fix these coefficients we construct the Gram matrix $G_{ij} = p_i \cdot p_j$ for $p_i^\mu \in \{v^\mu, k^\mu, q^\mu, n^\mu\}$. Explicitly, we find in four-dimensions
\begin{equation}
	G_{ij} = 
	\begin{pmatrix}
  		 1 & \omega & 0 & 0 \\[6pt]
    	\omega & -q^2/4 & 0 & 0 \\[6pt]
    	0 & 0 & q^2 & 0 \\[6pt]
    	0 & 0 & 0 & q^2(q^2 + 4 \omega^2)/4
	\end{pmatrix}\;, \quad 
	G_{ij}^{-1} = \frac{1}{q^2 + 4 \omega^2} 
	\begin{pmatrix}
  		 q^2 & 4\omega & 0 & 0 \\[6pt]
    	4\omega & -4 & 0 & 0 \\[6pt]
    	0 & 0 & (q^2+4\omega^2)/q^2 & 0 \\[6pt]
    	0 & 0 & 0 & 4/q^2
	\end{pmatrix}\;.
\end{equation}
Then, we obtain the coefficients \eqref{eq:ep1_decomp}, \eqref{eq:ep2_decomp} and \eqref{eq:a_decomp} using the inverse Gram matrix. In four-dimensions we find the following values
\begin{align}
	(q^2 + 4 \omega^2) \lambda_1 &  = 4 \omega (k \cdot \varepsilon_1) + q^2 (v \cdot \varepsilon_1 ) \;, \\
	(q^2 + 4 \omega^2) \lambda_2 &  = - 4 (k \cdot \varepsilon_1) + 4\omega (v \cdot \varepsilon_1 )\;, \\
	q^2 \lambda_3 &  = - 2 (k \cdot \varepsilon_1)\;, \\
	q^2(q^2 + 4\omega^2)\lambda_4 & =  4 (n \cdot \varepsilon_1)\;,
\end{align}
\vspace{-0.5cm}
\begin{align}
	(q^2 + 4 \omega^2) \xi_1 &  = 4 \omega (k \cdot \varepsilon_2) + q^2 (v \cdot \varepsilon_2 ) \;, \\
	(q^2 + 4 \omega^2) \xi_2 &  = - 4 (k \cdot \varepsilon_2) + 4\omega (v \cdot \varepsilon_2 )\;, \\
	q^2 \xi_3 &  = 2 (k \cdot \varepsilon_2)\;, \\
	q^2(q^2 + 4\omega^2)\xi_4 & =  4 (n \cdot \varepsilon_2)\;, 
\end{align}
\vspace{-0.5cm}
\begin{align}
	\hspace{-2cm}(q^2 + 4 \omega^2) \zeta_1 &  = 4 \omega (k \cdot a) \;, \\
	\hspace{-2cm}(q^2 + 4 \omega^2) \zeta_2 &  = - 4 (k \cdot a)\;, \\
	\hspace{-2cm}q^2 \zeta_3 &  = 2 (q \cdot a)\;, \\
	\hspace{-2cm}q^2(q^2 + 4\omega^2)\zeta_4 & =  4 (n \cdot a)\;.
\end{align}

\section{The Compton amplitude in a general frame}
\noindent To compare the non-minimal results with Kerr values of Wilson coefficients to the minimal results, it is convenient to choose an explicit frame. This will automatically impose four-dimensional identities. Without loss of generality, and satisfying the various on-shell constraints, we may choose the frame
\begin{align}
	& v^\mu = (1,0,0,0)\;, \quad a^\mu = (0,0,0, a)\;, \\
	& k_1^\mu = \omega r_1^\mu\;, \hspace{1.15cm} r_1^\mu = (1, \sin\theta_1 \cos \phi_1, \sin \theta_1 \sin \phi_1, \cos \theta_1)\;, \\
	& k_2^\mu = - \omega r_2^\mu \;, \hspace{0.9cm} r_2^\mu = (1, \sin\theta_2 \cos \phi_2, \sin \theta_2 \sin \phi_2, \cos \theta_2)\;,
\end{align}
with polarizations 
\begin{equation}
	\varepsilon_j^{\pm} = \frac{1}{\sqrt{2} }\left(\partial_{\theta_j}r_j^\mu \pm \frac{i}{\sin \phi_j} \partial_{\phi_j}r_j^\mu\right)\;.
\end{equation}
Note that here we have used the rotational freedom to align the spin vector with the $z$-axis. Evaluating the minimal four-point Compton amplitude in this frame with positive helicity yields
\begin{equation}
	\mathcal{M}^{(0)}_{4,\mathrm{cl}} (a ,\omega ;\theta_1, \phi_1; \theta_2, \phi_2) = \mathcal{M}^{(0)}_{s = 0}\sum_{i = 0}^{4} \frac{(\omega a)^n (\cos \theta_1 - \cos \theta_2)^n}{n!}\;,
\end{equation}
where 
\begin{equation}
	\mathcal{M}^{(0)}_{s = 0} = m^2 \kappa^2\frac{\cos[\frac{1}{2}(\theta_1 + \theta_2 + \phi_1 - \phi_2)] - \cos[\frac{1}{2}(\theta_1 + \theta_2 - \phi_1 + \phi_2)] + 2 i \cos[\frac{1}{2}(\phi_1 - \phi_2)] \cos[\frac{1}{2}(\theta_1 - \theta_2)]}{32(-1 + \cos \theta_1 \cos \theta_2 + \cos(\phi_1 - \phi_2) \sin \theta_1 \sin \theta_2)}\;.
\end{equation}
This yields a compact form of the tree-level amplitude for comparisons~\cite{Bautista:2021wfy, Bautista:2022wjf}. The same can be done with non-minimal values of Wilson coefficients, and even at the one-loop level. 
\section{Ranked Fourier integrals}
\noindent As discussed in the main text, to obtain the ranked Fourier integral we start with the scalar Fourier integral
\begin{align}
	I_\alpha(b^2) = \int \hat{\mathrm{d}}^4 q \ \hat{\delta}(2 m v \cdot q) \hat{\delta}( 2 k \cdot q) e^{i b \cdot q} (-q^2)^{-\alpha} = \frac{1}{m\omega}\frac{\Gamma(1-\alpha)}{2^{2\alpha + 2} \pi \Gamma(\alpha)} \frac{1}{|b|^{2-2\alpha}}\;,
\end{align}
and take derivatives with respect to $b^\mu$. Up to the rank $4$ integral, we find
\begin{align}
	& I_\alpha^\mu(b^2) = - i \frac{\partial I_\alpha(b^2)}{\partial b_\mu} = -\frac{i}{m\omega} \frac{\Gamma(2-\alpha)}{2^{2\alpha +1} \pi \Gamma(\alpha)} \frac{b^\mu}{|b|^{4-2\alpha}}\;, \\
	& I_\alpha^{\mu\nu}(b^2) = - i \frac{\partial I^{\mu}_\alpha(b^2)}{\partial b_\nu} = - \frac{1}{m\omega}\frac{\Gamma(2-\alpha)}{2^{2\alpha +1} \pi \Gamma(\alpha)}\frac{1}{|b|^{6-2\alpha}}\Big[|b|^2 \Pi^{\mu \nu} + (4-2\alpha) b^\mu b^\nu\Big]\;, \\
	& I_\alpha^{\mu\nu \gamma}(b^2) = - i \frac{\partial I^{\mu \nu}_\alpha(b^2)}{\partial b_\gamma} = \frac{i}{m\omega}\frac{\Gamma(2-\alpha)}{2^{2\alpha +1} \pi \Gamma(\alpha)}\frac{(4-2\alpha)}{|b|^{8-2\alpha}}\Big[|b|^2( b^\mu \Pi^{\nu \gamma} + b^\nu \Pi^{\gamma \mu} + b^\gamma \Pi^{\mu \nu}) + (6-2\alpha) b^\mu b^\nu b^\gamma\Big]\;, \\
	& I_\alpha^{\mu\nu \gamma \rho}(b^2) = - i \frac{\partial I^{\mu \nu \gamma}_\alpha(b^2)}{\partial b_\rho} = \frac{1}{m\omega}\frac{\Gamma(2-\alpha)}{2^{2\alpha +1} \pi \Gamma(\alpha)}\frac{(4-2\alpha)}{|b|^{10-2\alpha}}\Big[|b|^4( \Pi^{\mu \rho}\Pi^{\nu \gamma} + \Pi^{\nu \rho}\Pi^{\gamma \mu}+\Pi^{\gamma \rho}\Pi^{\mu \nu}) + (8-2\alpha)(6-2\alpha) b^\mu b^\nu b^\gamma b^\rho \nonumber \\ 
	& \hspace{4cm} + (6-2\alpha) |b|^2(b^\mu b^\rho \Pi^{\nu \gamma} + b^\nu b^\rho \Pi^{\gamma \mu} + b^\gamma b^\rho \Pi^{\mu \nu} + b^\mu b^\gamma \Pi^{\nu \rho} + b^\nu b^\gamma \Pi^{\mu \rho} + b^\mu b^\nu \Pi^{\gamma \rho}) \Big]\;.
\end{align}
In these expressions we replace all instances of the metric with the transverse projector
\begin{align}
	\Pi^{\mu \nu} = \eta^{\mu \nu} - \frac{v^\mu k^\nu + k^\mu v^\nu}{\omega} + \frac{k^\mu k^\nu}{\omega^2}\;,
\end{align}
which takes us to the plane orthogonal to $v^\mu$ and $k^\mu$.
\end{document}